\documentclass[traditabstract,longauth]{aa}
\usepackage{natbib}
\usepackage{epsfig}
\usepackage{txfonts}

\def\vsini{$v$\,sin\,$i$}             
\def\ms{\hbox{\,m\,s$^{-1}$}}         
\def\m2s2{\hbox{\,m$^{2}$\,s$^{-2}$}} 
\def\kms{\hbox{\,km\,s$^{-1}$}}       
\def\gcm3{\hbox{\,g\,cm$^{-3}$}}      
\def\vsini{\hbox{$v$\,sin\,$i$}}      
\def\Msun{\hbox{$M_{\odot}$}}         
\def\Rsun{\hbox{$R_{\odot}$}}

\begin{document}

\title{Transiting exoplanets from the CoRoT space mission\thanks{The
    CoRoT space mission, launched on December 27th 2006, has been
    developed and is operated by CNES, with the contribution of
    Austria, Belgium, Brazil , ESA (RSSD and Science Programme),
    Germany and Spain.}}

\subtitle{XIII. CoRoT-14b: an unusually dense very hot Jupiter }

\author{
B.~Tingley \inst{\ref{IAC},\ref{ULL}}
\and M.~Endl\inst{\ref{UT}}
\and J.-C.~Gazzano\inst{\ref{LAM}}
\and R.~Alonso\inst{\ref{Geneve}} 
\and T.~Mazeh\inst{\ref{Tel Aviv}} 
\and L.~Jorda\inst{\ref{LAM}} 
\and S.~Aigrain\inst{\ref{Oxford}} 
\and J.-M.~Almenara\inst{\ref{IAC},\ref{ULL}}
\and M.~Auvergne\inst{\ref{LESIA}} 
\and A.~Baglin\inst{\ref{LESIA}}
\and P.~Barge\inst{\ref{LAM}} 
\and A.~S.~Bonomo\inst{\ref{LAM}} 
\and P.~Bord\'e\inst{\ref{IAS}} 
\and F.~Bouchy\inst{\ref{OHP},\ref{IAP}} 
\and H.~Bruntt\inst{\ref{LESIA}}
\and J.~Cabrera\inst{\ref{DLR},\ref{LUTh}} 
\and S.~Carpano\inst{\ref{ESA}}
\and L.~Carone\inst{\ref{Koeln}} 
\and W.~D.~Cochran\inst{\ref{UT}}
\and Sz.~Csizmadia\inst{\ref{DLR}} 
\and M.~Deleuil\inst{\ref{LAM}} 
\and H.~J.~Deeg\inst{\ref{IAC},\ref{ULL}} 
\and R.~Dvorak\inst{\ref{Wien}} 
\and A.~Erikson\inst{\ref{DLR}}
\and S.~Ferraz-Mello\inst{\ref{Brasil}} 
\and M.~Fridlund\inst{\ref{ESA}}
\and D.~Gandolfi\inst{\ref{Tautenburg}} 
\and M.~Gillon\inst{\ref{Geneve},\ref{Liege}} 
\and E.~W.~Guenther\inst{\ref{Tautenburg}} 
\and T.~Guillot\inst{\ref{OCA}} 
\and A.~Hatzes\inst{\ref{Tautenburg}} 
\and G.~H\'ebrard\inst{\ref{IAP}} 
\and A.~L\'eger\inst{\ref{IAS}} 
\and A.~Llebaria\inst{\ref{LAM}} 
\and H.~Lammer\inst{\ref{Graz}} 
\and C.~Lovis\inst{\ref{Geneve}}
\and P.~J.~MacQueen\inst{\ref{UT}}
\and C.~Moutou\inst{\ref{LAM}} 
\and M.~Ollivier\inst{\ref{IAS}} 
\and A. Ofir\inst{\ref{Wise}}
\and M.~P\"atzold\inst{\ref{Koeln}} 
\and F.~Pepe\inst{\ref{Geneve}}
\and D.~Queloz\inst{\ref{Geneve}}
\and H.~Rauer\inst{\ref{DLR},\ref{ZAA}} 
\and D.~Rouan\inst{\ref{LESIA}}
\and B.~Samuel\inst{\ref{IAS}}
\and J.~Schneider\inst{\ref{LUTh}} 
\and A.~Shporer\inst{\ref{Wise}}
\and G.~Wuchterl\inst{\ref{Tautenburg}} 
}

\institute{
Instituto de Astrof\'{i}sica de Canarias, E-38205 La Laguna, Tenerife, Spain\label{IAC}
\and Dpto. de Astrof\'isica, Universidad de La Laguna, 38206 La Laguna, Tenerife, Spain\label{ULL}
\and McDonald Observatory, University of Texas at Austin, Austin, 78712 TX, USA\label{UT}
\and Laboratoire d'Astrophysique de Marseille, 38 rue Fr\'ed\'eric Joliot-Curie, 13388 Marseille cedex 13, France\label{LAM}
\and Observatoire de l'Universit\'e de Gen\`eve, 51 chemin des Maillettes, 1290 Sauverny, Switzerland\label{Geneve}
\and Wise Observatory, Tel Aviv University, Tel Aviv 69978, Israel\label{Wise}
\and Department of Physics, Denys Wilkinson Building Keble Road, Oxford, OX1 3RH\label{Oxford}
\and LESIA, UMR 8109 CNRS, Observatoire de Paris, UPMC, Universit\'{e} Paris-Diderot, 5 place J. Janssen, 92195 Meudon, France\label{LESIA}
\and Institut d'Astrophysique Spatiale, Universit\'e Paris-Sud 11 \& CNRS (UMR 8617), B\^{a}t. 121, 91405 Orsay, France\label{IAS}
\and Observatoire de Haute Provence, 04670 Saint Michel l'Observatoire, France\label{OHP}
\and Institut d'Astrophysique de Paris, 98bis boulevard Arago, 75014 Paris, France\label{IAP}
\and University of Vienna, Institute of Astronomy, T\"urkenschanzstr. 17, A-1180 Vienna, Austria\label{Wien}
\and Institute of Planetary Research, German Aerospace Center, Rutherfordstrasse 2, 12489 Berlin, Germany\label{DLR}
\and IAG, Universidade de Sao Paulo, Brazil\label{Brasil}
\and Research and Scientific Support Department, ESTEC/ESA, PO Box 299, 2200 AG Noordwijk, The Netherlands\label{ESA} 
\and University of Li\`ege, All\'ee du 6 ao\^ut 17, Sart Tilman, Li\`ege 1, Belgium\label{Liege}
\and Th\"uringer Landessternwarte, Sternwarte 5, Tautenburg 5, D-07778 Tautenburg, Germany\label{Tautenburg}
\and Space Research Institute, Austrian Academy of Science, Schmiedlstr. 6, A-8042 Graz, Austria\label{Graz}
\and School of Physics and Astronomy, Raymond and Beverly Sackler Faculty of Exact Sciences, Tel Aviv University, Tel Aviv, Israel\label{Tel Aviv}
\and Rheinisches Institut f\"ur Umweltforschung an der Universit\"at zu K\"oln, Aachener Strasse 209, 50931, Germany\label{Koeln}
\and Observatoire de la C\^ote d'Azur, Laboratoire Cassiop\'ee, BP 4229, 06304 Nice Cedex 4, France\label{OCA}
\and LUTH, Observatoire de Paris, CNRS, Universit\'e Paris Diderot; 5 place Jules Janssen, 92195 Meudon, France\label{LUTh}
\and Center for Astronomy and Astrophysics, TU Berlin, Hardenbergstr. 36, 10623 Berlin, Germany\label{ZAA}
\and School of Physics, University of Exeter, Stocker Road, Exeter EX4 4QL, United Kingdom\label{Exeter}
\and Laboratoire d'Astronomie de Lille, Universit\'e de Lille 1, 1 impasse de l'Observatoire, 59000 Lille, France\label{Lille}
\and Institut de M\'ecanique C\'eleste et de Calcul des Eph\'em\'erides, UMR 8028 du CNRS, 77 avenue Denfert-Rochereau, 75014 Paris, France\label{IMCCE}
}
\date{Received ; accepted }

\abstract{In this paper, the CoRoT Exoplanet Science Team announces
  its 14th discovery. Herein, we discuss the observations and analyses
  that allowed us to derive the parameters of this system: a hot
  Jupiter with a mass of $7.6 \pm 0.6$ Jupiter masses orbiting a
  solar-type star (F9V) with a period of only 1.5 d, less than 5
  stellar radii from its parent star. It is unusual for such a massive
  planet to have such a small orbit: only one other known exoplanet
  with a higher mass orbits with a shorter period.

}

\titlerunning{CoRoT-14b}
\authorrunning{Tingley et al.}

\maketitle

%
\section{Introduction}

Transiting exoplanets offer greater opportunities for the study and
understanding of exoplanetary systems than those discovered by radial
velocity measurements. Analysis of transit light curves yields
planetary radii and enables tests for rings \citep{barnes2004}, moons
\citep{sart1999}, and other planets through transit timing variations
\citep{mac2010}, while high-precision observations of primary and
secondary transits can reveal some details of planetary atmospheres
(which is not currently possible for non-transiting planets) and
albedos \citep[e.g.]{dem2009}, which is easier for transiting
exoplanets but still possible for others.

The potential of transiting exoplanets has inspired considerable
effort towards their discovery, both from the ground and from
space. While ground-based searches have discovered the majority of
known transiting exoplanets to this point, space-based missions offer
the greatest potential for discovery. Observing from space allows
nearly continuous sampling and much better photometric precision,
which is adversely affected by the atmosphere. This makes it possible
to detect long-period transiting exoplanets, whose transits can easily
be longer than a typical night, and smaller exoplanets, whose transits
are too shallow to be detected from ground.

The CoRoT (CO{\it vection} RO{\it tation and planetary }T{\it ransits}
space mission was the first space mission dedicated primarily to
searching for transits \citep{bag2009}. The mission has successfully
demonstrated the advantages to space; given its orbit and the lack of
atmosphere, it can observe the same field continuously for up to five
months with remarkably high relative precision. This enabled the
discovery of both the first transiting 'Super-Earth'
\citep[CoRoT-7b:]{leg2009,que2009} and the first temperate transiting
gas giant \citep[CoRoT-9b]{dee2010}.

In this paper, we announce the discovery of the 14th transiting planet
discovered by CoRoT; an unusually massive exoplanet orbiting an F9V
star with metallicity consistent with Solar. In Sec. 2, we detail the
CoRoT photometry. In Sec. 3, we describe the ground-based follow-up
observations that we used to confirm the planetary nature of
CoRoT-14b. In Sec. 4, we discuss our analysis of the light curves to
extract the transit parameters and present the inferred planetary
parameters.  In Sec. 5, we analyze the parent star. Finally, we
conclude our paper in Sec. 6, where we discuss how the properties
CoRoT-14b compare to the ensemble of known transiting planets.

%
\section{CoRoT observations}

CoRoT monitored the field which contains CoRoT-14b during its second
Long-Run Anti-center pointing ({\it LRa02}). This run lasted from 16
Nov 2008 to 11 Mar 2009 and images a 3.5 square degree
field in the constellation {\it Monoceros}. The details of the
observations that comprise this run will appear in a forthcoming paper.
Table 1 lists various IDs, coordinates, and magnitudes for CoRoT-14b.

\begin{figure}[h]
\centering
\includegraphics[width=9cm]{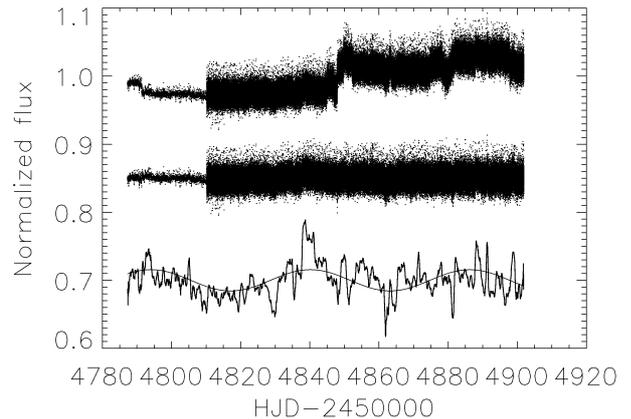}
\caption{The processed and normalized transit light curve of
  CoRoT-14. The top plot shows the final processed photometry which
  corrects for jitter and other effects, but not hot pixels -- which
  clearly have a strong effect. The middle plot shows the same data,
  corrected for hot pixels. These {\bf plots} show that the integration time
  changed from 512s to 32s at HJD = 2454810.2. The bottom plot is the
  smoothed light curve, multiplied by a factor of ten to show
  detail. It emphasizes the low level of activity, around 2 mmag. See
  Sect. 5 for more information on the activity of the parent
  star. While the light curve on the surface appears to be slightly
  sinusoidal, this is in fact not the case: removing the best-fit
  sinusoid (which has an amplitude of about 1.6 mmag and a period of
  about 46 days, corresponding to neither the rotation period of the
  star, the period of the transit, nor any known instrumental effects)
  reduces the $\sigma_{\rm rms}$ by only about 10\%.}
\label{photometry}
\end{figure}

CoRoT-14b was first identified as an object of interest on 9 Dec 2008
by the 'alarm mode' pipeline \citep{sur2008} and the
time sampling subsequently switched from the standard value of $512s$
to the $32s$ sampling reserved for interesting
targets. Figure~\ref{photometry} shows the final monochromatic
light curve, containing 220188 photometric samples covering over 114
days. This light curve is the output of the standard CoRoT pipeline
\citep[version 2.1, see]{auv2009} in conjunction
with further processing to remove outliers and correct for
systematics, as described in, e.g. \citet{bar2008} and
\citet{alo2008}.  It exhibits many discontinuities due
to cosmic ray hits on the detector -- a common occurrence in CoRoT
light curves, as the satellite passes through the
energetic-particle-rich South-Atlantic Anomaly each orbit. These can
be corrected for, however, yielding a fairly
good cleaned light curve with a $\sigma_{\rm rms}$ of only around
2 mmag, indicating that the star is not particularly active.

The periodic transit signals are easily detectable in the cleaned
light curve. It contains some 89 transits, 74 of them after the
sampling rate increased, yielding a final duty cycle of 82.7\%. The
initial trapezoid fitting, using the method outlined in
\citet{alo2008} yielded a period ($P$) of $1.15214 \pm 0.00013d$ and a
      {\bf primary transit epoch} ($T_0$) of $2454787.6694 \pm
      0.00053$ and a depth of about 5 mmag. This information was
      passed on to the photometric and spectroscopic follow-up groups,
      to help schedule the ground-based follow-up observations
      necessary for confirmation or rejection.

\begin{table}
\caption{IDs, coordinates and magnitudes.}
\centering
\begin{tabular}{lcc}       
\hline\hline                 
CoRoT window ID & LRa02 E2 5503 \\
CoRoT ID        & 110864907 \\
USNO-A2 ID      & 0825-03434910 \\
2MASS ID        & J06534181-0532097 \\
GSC2.3 ID       & S10023017631 \\
\\
\multicolumn{2}{l}{Coordinates} \\
\hline            
RA (J2000)  &  06:53:41.80 \\
Dec (J2000) & -05:32:09.82 \\
\\
\multicolumn{3}{l}{Magnitudes} \\
\hline
\centering
Filter & Mag & Error \\
\hline
B\tablefootmark{a}  & 16.891 & 0.156\\
V\tablefootmark{a}  & 16.033 & 0.070\\
r'\tablefootmark{a} & 15.672 & 0.048\\
J\tablefootmark{b}  & 14.321 & 0.036 \\
H\tablefootmark{b}  & 14.007 & 0.047 \\
K\tablefootmark{b}  & 13.806 & 0.053 \\

\hline
\end{tabular}
\tablefoot{
  \tablefoottext{a}{Provided by Exo-Dat (Deleuil et al, 2009) based on 4-color photometry taken at the 2.5m INT.}
  \tablefoottext{b}{from 2MASS catalog.}
}
\label{startable}      
\end{table}

\section{Ground-based observations}

The detection of a transit-like event in a light curve is only the
beginning of the process: we find 10 to 20 candidates with CoRoT for
each planet. In order to exclude as many candidates as possible
without resorting to precious HARPS/HIRES observing time, we perform a
carefully considered sequence of ground-based follow-up observations.

\subsection{Imaging - contamination}

The first step is to image the field around the start for possible
sources of photometric contamination that may combine with light from
the target star to masquerade as a transit-like event and to estimate
how much, if at all, nearby stars dilute the transit
\citep{dee2009}. This is necessary for CoRoT in particular because the
light passes through bi-prism to disperse the light over many
pixels. While this allows much longer exposures (much like an ordinary
defocusing would have) and some color information, it comes at the
expense of an increase in contamination from nearby (and occasionally
not so nearby) stars.

The photometric follow-up group obtained 20 images of the candidate
during mid-transit on 27 February 2009 and 20 images out-of-transit on
14 April 2009 with the IAC80 telescope on Tenerife, which has an
aperture of 80 cm and a 10.6 x 10.6 arcminute field of view. Analysis
revealed no strong signals in nearby stars that could be capable of
producing a false positive nor any bright very close neighbors, but
was not sensitive enough to detect the transit with any
confidence. {\bf These images, when stacked, are of similar quality and
depth to those in large surveys such as 2MASS, so we are confident
that no unknown, readily identifiable stars have eluded us.}
Contamination analysis \citep{dee2009} revealed that $93
\pm 0.5\%$ of total light in the CoRoT mask came from target, with
most of the rest ($\sim6\%$) coming from a star about 3 magnitudes
fainter and 2.5 arcseconds to the south. This factor was included in
the final transit analysis in Sec.~\ref{transitsec}.

\subsection{Radial velocities - spectroscopy / orbital fit}

We planned radial velocity (RV) observations only after the
photometric imaging with the IAC80 showed that this candidate had but
a slight risk of being a false positive.  Radial velocity (RV)
observations of CoRoT-14 were performed with the HARPS spectrograph
\citep{mayor03}, based on the 3.6\,m ESO telescope (Chile) and the
HIRES spectrograph \citep{vogt94} installed on the 10\,m
Keck telescope in Hawaii.

HARPS was used with the observing mode obj\_AB, without simultaneous
thorium in order to monitor the Moon background light on the second
fiber.  The intrinsic stability of this spectrograph frees us from the
need to capture a simultaneous thorium spectrum, as the instrumental
drift during an exposure is always smaller than the stellar RV photon
noise uncertainties in this case. We took a series of 8 spectra with
one hour exposures between November 22th 2009 and February 20th 2010
(ESO program 184.C-0639). We analyzed the HARPS data with the pipeline
based on the cross-correlation techniques
\citep{baranne96,pepe02}. The signal-to-noise per pixel at 550 nm of
individual spectrum is in the range 3.7 to 7.1 for this faint target,
one of the faintest followed-up by HARPS.  Radial velocities were
obtained by weighted cross-correlation with a numerical G2 mask.

We used HIRES in combination with its iodine (I$_2$) cell to measure
precise RVs. All observations were taken with a 7~arcsec long slit of
0.861~arcsec width, which yields a spectral resolving power of
$R\approx50,000$.  We obtained one spectrum of CoRoT-14 on 2009
December 5th without the I$_2$-cell to serve as stellar template for
the RV computation, which is required for calibration, and to
determine stellar parameters.  We took a single 1200 second exposure,
which had a signal-to-noise ratio (per pixel) of only 10 at 550\,nm,
as seeing conditions on this particular night were less than
optimal. We also took one exposure with the I$_2$-cell, to get an RV
measurement that night. We collected five additional spectra of
CoRoT-14 during January 2010 with the I$_2$-cell over the course of
three nights. The signal-to-noise ratios of these data range from 15
to 19 (at 550\,nm). We used our {\it Austral} Doppler code
\citep{endl00} to compute precise differential RVs. The results are
given in Table~\ref{rvtable}. Since the template spectrum had such
poor S/N, we used a HIRES template of a similar, but much brighter,
star (HD~12800) for the RV computation.

The results of the bisector analysis accompany the corresponding
radial velocity measurements in Table~\ref{rvtable}. The bisector
analysis was only possible with the HARPS spectra; the HIRES spectra
do not have sufficient resolution to yield meaningful results in this
case. The bisectors weakly anti-correlate with the differential RVs
(linear correlation coefficient $R = -0.198$), which in turn yields a
probability of 0.637 that the bisectors and RVs are physically {\em
  unrelated}. We therefore conclude that the bisector analysis is
consistent, albeit weakly, with no correlation.

\begin{figure}[h]
\centering
\includegraphics[width=6.5cm, angle=-90]{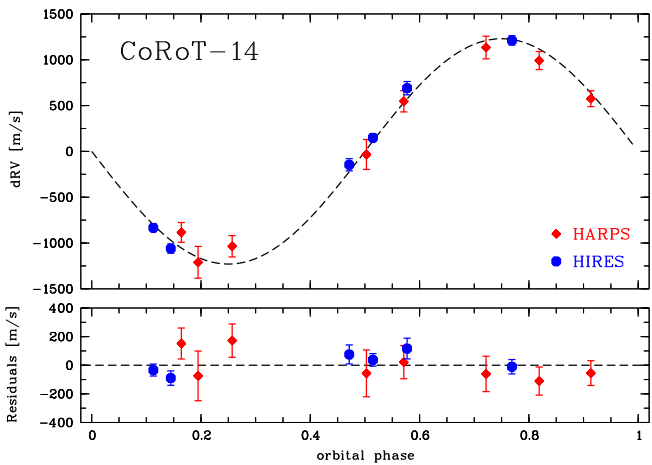}
\caption{RV orbital fit. This figure shows the orbital fit to
the HIRES and HARPS observations, using the period found by the
CoRoT photometry, along with the residuals, assuming a circular
orbit. A fit was made without this assumption, but returned a
value for the eccentricity that was consistent with zero.}
\label{orbit}
\end{figure}

We computed a Keplerian orbital solution for the HARPS and HIRES RV
data using the {\it Gaussfit} generalized least-squares software of
\citet{jeff88}.  We kept the values of the orbital period and {\bf
  primary transit epoch} fixed to the parameters determined by the
CoRoT photometry.  The individual velocity zero-points of the HARPS
and HIRES data were included as free parameters in the fitting
process. We first fit a circular orbit to the data (see
Figure~\ref{orbit}).  The $\chi^{2}_{\rm red}$ of this solution is
1.50 and the values of the residual rms scatter around the fit are
118\,m\,s$^{-1}$ (HARPS) and 78\,m\,s$^{-1}$ (HIRES). The orbital fit
yields an RV semi-amplitude $K$ of $1230\pm34$\,m\,s$^{-1}$. Adopting
a stellar mass of $1.13\pm0.09$\,M$_{\odot}$ for CoRoT-14 (see next
section), we obtain a mass of $7.6\pm0.6$\,M$_{\rm Jup}$ for the
planet. From this, we can conclude that CoRoT-14b is very massive gas
giant for its relatively short 1.5\,day orbit and orbits only
0.027\,AU from its parent star. The orbital parameters are summarized
in Table~\ref{starplanet_param_table}.

We also explored the possibility of an eccentric orbit. Allowing
eccentricity and periastron argument to be free parameters, we derive
an eccentricity of $e=0.019\pm0.046$, which a $\chi^{2}_{\rm red}$ of
1.63.  We therefore conclude that the current RV data for CoRoT-14 are
consistent with a completely circularized orbit.

\begin{table}[h]
 \caption{Radial velocity measures, errors, and bisectors. 
} \centering
 \begin{minipage}[!]{7.0cm}  
 \begin{tabular}{lccc}       
 \hline
 \hline                
BJD & RV & $\sigma_{RV}$ & Bisector \\
(days) & (\kms) & (\kms) & (\kms) \\
\hline
HARPS & & & \\
\hline
24555157.72444 & 7.8797 & 0.1237 & -0.3163 \\
24555235.65345 & 5.7109 & 0.1164 & 0.3223 \\
24555237.64047 & 7.2927 & 0.1159 & 0.1372 \\
24555239.67027 & 7.3209 & 0.0866 & -0.3827 \\
24555244.58565 & 5.8620 & 0.1081 & -0.0784 \\
24555245.57587 & 7.7376 & 0.0982 & -0.0693 \\
24555246.60996 & 6.7125 & 0.1636 & 0.1157 \\
24555247.65632 & 5.5354 & 0.1734 & -0.3999 \\
\hline
\hline
BJD & RV & $\sigma_{RV}$ \\
(days) & (\ms) & (\ms) \\
\hline
HIRES & &  \\
\hline
24555170.9552 &  -69.0 & 66.5 \\
24555221.8744 & -983.6 & 50.4 \\
24555222.8182 & 1288.0 & 50.4 \\
24555223.9457 &  226.3 & 44.0 \\
24555224.0403 &  766.7 & 72.2 \\
24555224.8502 & -757.0 & 41.7 \\

\hline
\end{tabular}
\end{minipage}
\label{rvtable}
\end{table}

\section{Analysis of the transit\label{transitsec}}

\begin{figure}[h]
\centering
\includegraphics[width=9cm]{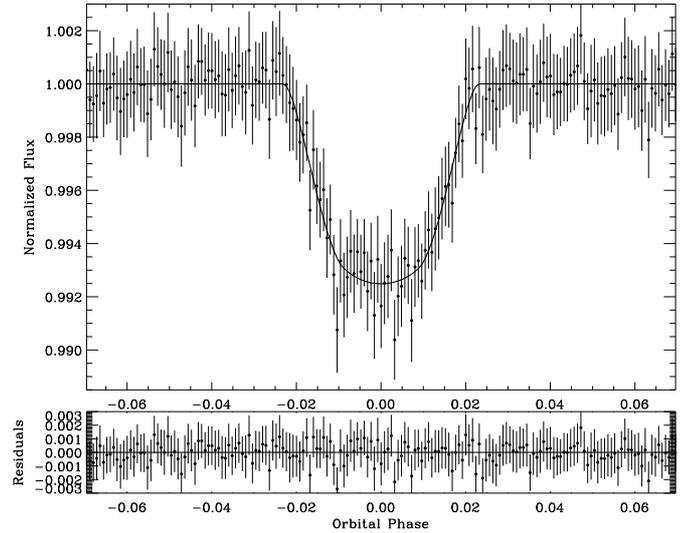}
\caption{Phase-folded transit and residuals. This figure shows
the phase-folded transit from the CoRoT photometry with the best-fit
model transit (top) along with the O-C residuals (bottom).}
\label{transit}
\end{figure}

We use the methodology described in \citet{alo2008} to
extract the transit parameters from the CoRoT photometry. To
summarize: we use trapezoid fitting to obtain the period and transit
epoch, then use a $\chi^2$ analysis described by
\citet{gim2006} on the phase-folded transit to determine
transit and stellar parameters (the transit center $T_c$, the orbital
phase at first contact $\theta_1$, the ratio of radii $k$, the orbital
inclination $i$ and $u_+$ and $u_-$ coefficients, which are related to
the quadratic limb darkening coefficients. We performed the transit
fitting using a bootstrap analysis to constrain parameters space,
based on the prescription outlined in \citet{bar2008} and
\citet{alo2008}. Due to the faintness of the target, we
chose not to fit the limb darkening coefficients: instead, we took the
values from \citet{sing10}, with (conservative) error bars for
these based on the uncertainties in the stellar parameters ($u_a=0.43
\pm 0.03$ and $u_b=0.24 \pm 0.1$). For each of the 500 bootstrap
curves, we fixed the limb darkening coefficients, but instead of
always using the same values, we extracted them from an random normal
distribution with the appropriate width. Thus, for each bootstrap
curve, we changed the contamination factor, the limb darkening
coefficients, and the residuals, which we shifted circularly, with the
initial parameters for the amoeba minimization algorithm perturbed
randomly.  The results of this analysis can be found in
Table~\ref{starplanet_param_table} and the transit and best fit can be
seen in Figure~\ref{transit}.

\section{Analysis of the parent star}

\begin{figure}
\centering
\includegraphics[width=3.25cm, angle=90]{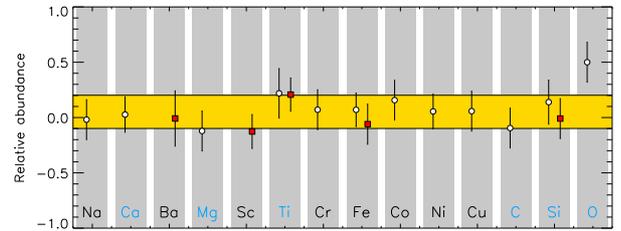}
\caption{Mean abundances for 14 elements in CoRoT-14 \emph{HARPS}
  spectrum. White circles correspond to neutral lines, red boxes to
  singly ionized lines and the yellow area represents the mean
  metallicity within one sigma error bar.}
\label{fig:ab}
\end{figure}

We co-added the \emph{HARPS} spectra to perform the analysis of the
parent star, which yielded R~$\sim$~110\,000 and $S/N\simeq45$ at
5500~\AA. From this, we were able to determine the $v\sin i$
($=9\pm0.5$~km\,s$^{-1}$). We obtained a first estimation of the
effective temperature of $\sim 5900$~K by fitting the H$_{\alpha}$
line.  We used these values as a starting point for the detailed
analysis of the \emph{HARPS} spectra with the \emph{VWA} package
\citep{bruntt10}. This analysis returned the following atmospheric
parameters: T$_{\rm eff}= 6035\pm100$~K, $\log\,g=4.35\pm0.15$~cgs,
[M/H]$=0.05\pm0.15$~dex, plus individual abundances for several
elements, which are listed in Table \ref{tab:ab} and shown in
figure~\ref{fig:ab}.

\begin{table}[!h]
 \centering
 \caption{Abundances of some chemical elements for the fitted lines in
   the \emph{HARPS} spectrum. The listed abundances are relative to
   the solar value.  Last column gives the number of lines used.
 \label{tab:ab}}
 \setlength{\tabcolsep}{3pt} 
\begin{tabular}{lrr}
\hline
\hline
\noalign{\smallskip}
Element & [X/H] (1$\sigma$) & Nb Lines  \\
\noalign{\smallskip}
\hline
\noalign{\smallskip}
 {Ca \sc   i} &    0.03  (0.16)  &   8 \\
 {Ba \sc  ii} &   $-$0.01  (0.25)  &   4 \\
 {Sc \sc  ii} &  $-$0.13  (0.15)  &   4 \\
 {Ti \sc   i} &   0.22  (0.22)  &   5 \\
 {Ti \sc  ii} &    0.21  (0.15)  &   8 \\
 {Fe \sc   i} &    0.05  (0.15)  &  38 \\
 {Fe \sc  ii} &   $-$0.06  (0.18)  &   6 \\
 {Ni \sc   i} &    0.06  (0.16)  &  21 \\
 {Si \sc   i} &    0.14  (0.20)  &   4 \\
 {Si \sc  ii} &   -0.01  (2.01)  &   1 \\
 \noalign{\smallskip}
\hline

\end{tabular}
\end{table}

The large error bars, especially on the surface gravity and the
metallicity, are due to the low signal-to-noise ratio of the spectra
due to the faintness of the star. Using the density from the transit
fit and the effective temperature and the metallicity from the
spectroscopic analysis, we derived a mass of $1.13\pm0.09M_\odot$, and
a radius of $1.21\pm0.08R_\odot$ for the star using the dedicated
\emph{STAREVOL} evolutionary tracks \citep{PalaciosPC,Siess06}. As a
final check, we ascertained that the inferred surface gravity agreed
with the spectroscopic value, $\log\,g_{\rm evol}= 4.33\pm0.14 $~cgs.

\begin{figure}[h]
\centering
\includegraphics[width=6.5cm, angle=270]{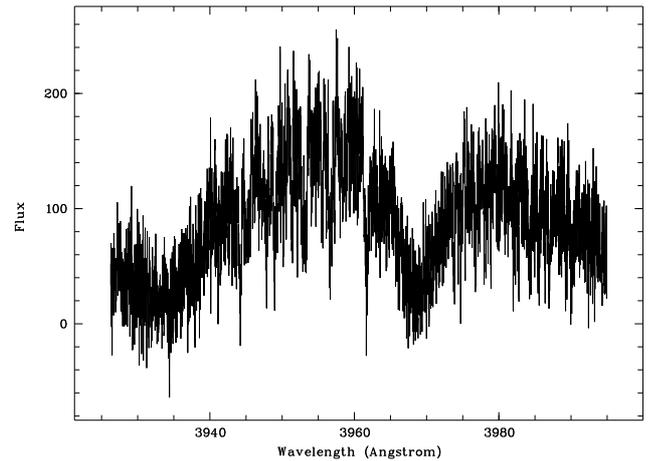}
\caption{Calcium II H and K lines. This plot shows the
Calcium II H and K lines obtained from the HIRES
spectra, with Ca II K on the left (centered at about 3929\AA)
and Ca II H on the right (centered at about 3978\AA. No
evidence for emission in the cores of these lines can be seen,
which is consistent with a star of low magnetic activity.}
\label{caiihk}
\end{figure}
\begin{figure}[h]
\centering
\includegraphics[width=8.5cm]{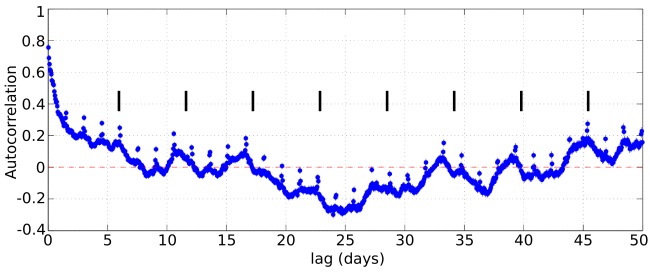}
\caption{Rotational Period from Auto-correlation. This figure shows
  the auto-correlation of the CoRoT-14 photometry, which is created
  by correlated the light curve with a temporally-shifted version of
  itself (the lag listed on the X-axis). The broad local maxima at
  multiples of 5.66 days (marked by the vertical black lines)
  correspond to the rotation of the star, visible through the
  photometric footprints of activity-induced variations on the stellar
  photosphere. Also apparent in this figure are number short, sharp
  periodic features: these are the periodic transit that the
  autocorrelation function detects.}
\label{rotation}
\end{figure}
\begin{figure}[h]
\centering
\includegraphics[width=8.5cm]{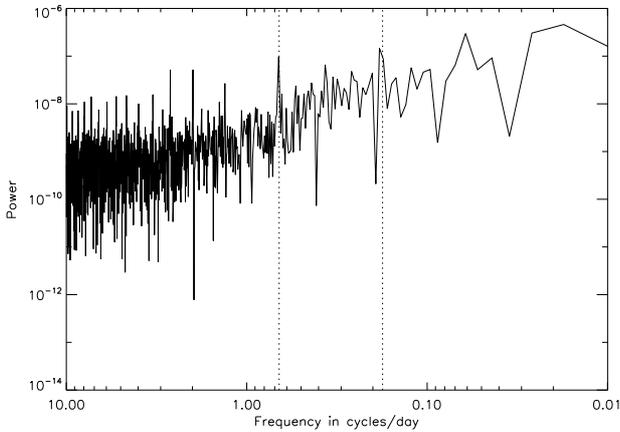}
\caption{Rotational Period using the Discrete Fourier Transform. This
  figure shows the discrete Fourier transform power series of the
  CoRoT-14 photometry. This approach also detects the planetary
  transit and confirms the rotation period (depicted by dotted lines)
  found by the auto-correlation, albeit less convincingly.}
\label{rotation2}
\end{figure}

As the RV spectra are slightly less than ideal, an examination of the
activity of the parent star is warranted. While the star is
photometrically variable on the level of only a few millimags, other
methods can be use to corroborate this, in particular the Ca II H and
K lines. These are shown in figure~\ref{caiihk} and show no evidence
for emission in the cores of these lines, which is consistent with a
star of low magnetic activity. While the activity level is low, it is
non-zero. We decided therefore to attempt to estimate the stellar
rotation period from the \emph{CoRoT} light curve using an
auto-correlation analysis. The results of this analysis can be seen in
figure~\ref{rotation}. We discover local peaks in the auto-correlation
that are separated by 5.66 days, which we infer to be the rotation
period of the star. This compares fairly well with the rotational
period that can be inferred from $v\sin i$ and the radius of the star,
would be $6.8 \pm 0.8$ days, assuming the stellar rotation axis is
perpendicular to the line of sight. {\bf This result is confirmed by a
  discrete Fourier transform of the photometric time series, although
  the results are somewhat less convincing (see
  figure~\ref{rotation2}).}


{\bf We estimated the distance of the star to be $1340\pm110$ pc by
comparing the T$_{\rm eff}$ to the tables in Allen's Astrophysical
Quantities \citep{cox00} to obtain the absolute V magnitude and
corresponding (J-K) color to constrain the extinction. This was then
combined with the observed V magnitude to get the distance.}

\begin{table}
\caption{Planet and star parameters.}   
\begin{tabular}{l l}
\hline\hline                 
\multicolumn{2}{l}{\emph{Ephemeris}} \\
\hline
Planet orbital period $P$ [days]                         &  1.51214 $\pm$ 0.00013 \\
Primary transit epoch $T_\mathrm{tr}$ [HJD-2\,400\,000]  &  54787.6694 $\pm$ 0.0053 \\
Primary transit duration $d_\mathrm{tr}$ [h]             &  1.662 $\pm$ 0.044 \\
& \\
\multicolumn{2}{l}{\emph{Results from radial velocity observations}} \\
\hline    
Epoch of periastron $T_{0}$ [HJD-2\,400\,000] &  $54787.6702$ (fixed)  \\
Orbital eccentricity $e$                      &  $0$ (fixed)  \\
Radial velocity semi-amplitude $K$ [\ms]      &  $1230 \pm 34$  \\
& \\
\multicolumn{2}{l}{\emph{Fitted transit parameters}} \\
\hline
Radius ratio $k=R_\mathrm{p}/R_{*}$                    & 0.0925 $\pm$ 0.0019 \\
Limb darkening coefficients\tablefootmark{a} $u_+=u_a+u_b$ &  0.67 $\pm$ 0.03 \\
                                                     $u_-=u_a-u_b$ &  0.19 $\pm$ 0.03 \\
Inclination $i$ [deg]                              & 79.6 $\pm$ 0.8 \\

& \\
\multicolumn{2}{l}{\emph{Deduced transit parameters}} \\
\hline
Scaled semi-major axis $a/R_{*}$                   & 4.78 $\pm$ 0.28 \\
$M^{1/3}_{*}/R_{*}$ [solar units]                  & 0.86 $\pm$ 0.02\\
Stellar density $\rho_{*}$ [\gcm3]                 & 0.91 $\pm$ 0.17 \\
Impact parameter\tablefootmark{b} $b$                  & 0.86 $\pm$ 0.02 \\
& \\
\multicolumn{2}{l}{\emph{Spectroscopic parameters }} \\
\hline
Effective temperature $T_\mathrm{eff}$ [K]  & 6035 $\pm$ 100 \\
Surface gravity log\,$g$ [dex]              & 4.35 $\pm$ 0.15 \\
Metallicity $[\rm{Fe/H}]$ [dex]             & 0.05 $\pm$ 0.15 \\
Stellar rotational velocity {\vsini} [\kms] & 9.0 $\pm$ 0.5 \\
Spectral type                               & F9V        \\
& \\
\multicolumn{2}{l}{\emph{Stellar and planetary physical parameters from combined analysis}} \\
\hline
Star mass [\Msun]                                    & 1.13 $\pm$ 0.09 \\
Star radius [\Rsun]                                  & 1.21 $\pm$ 0.08 \\
Distance of the system [pc]                          & 1340 $\pm$ 110 \\
Stellar rotation period $P_\mathrm{rot}$ [days]      & 5.7 \\
Age of the star $t$ [Gyr]                            & 0.4 - 8.0 \\
Orbital semi-major axis $a$ [AU]                     & 0.0270 $\pm$ 0.002 \\
Planet mass $M_\mathrm{p}$ [M$_\mathrm{J}$ ]\tablefootmark{c}  &  $ 7.6 \pm 0.6$ \\
Planet radius $R_\mathrm{p}$ [R$_\mathrm{J}$]\tablefootmark{c} &  $ 1.09 \pm 0.07$ \\
Planet density $\rho_\mathrm{p}$ [$g\;cm^{-3}$]                &  $ 7.3 \pm 1.5$ \\
Equilibrium temperature\tablefootmark{d} $T^\mathrm{per}_\mathrm{eq}$ [K] &  $1952 \pm 66$ \\
\hline       
\end{tabular}
\tablefoot{ \tablefoottext{a}{$I(\mu)/I(1)=1-u_a(1-\mu)-u_b(1-\mu)^2$,
    where $I(1)$ is the specific intensity at the center of the disk and
    $\mu=\cos{\gamma}$, where $\gamma$ is the angle between the
    surface normal and the line of sight.}
  \tablefoottext{b}{$b=\frac{a \cdot \cos{i}}{R_{*}}$}
  \tablefoottext{c}{Radius and mass of Jupiter taken as 71492 km and 1.8986$\times$10$^{30}$ g, respectively.}
  \tablefoottext{d}{Zero albedo equilibrium temperature for an isotropic planetary emission.}
}
\label{starplanet_param_table}  
\end{table}

\section{Discussion}

\begin{figure}[h]
\centering
\includegraphics[width=8.5cm]{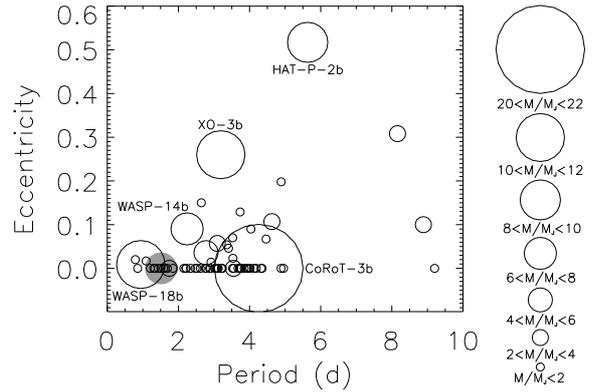}
\caption{Planetary period vs. eccentricity and mass. This figure shows
  period plotted against eccentricity for all transiting exoplanets
  with periods less than 10 days, with circle size indicative of
  planetary mass. Two-thirds of the 12 planets with $M > 2M_J$ have
  non-zero eccentricities; those with zero eccentricity include the
  lowest mass planet in the sample (Kepler-5b) and the most massive
  (CoRoT-3b). CoRoT-14b stands out by virtue of its high mass and
  short period -- only WASP-18b has a higher mass and a shorter
  period. Interestingly, WASP-18b has a small but significantly
  non-zero eccentricity \citep{triaud2010}. All
  planetary parameters from the Exoplanet Encyclopedia
  (http://exoplanet.eu), except for CoRoT-14b.}
\label{per_ecc_mass}
\end{figure}

The most interesting quality of CoRoT-14b is its mass relative to its
period -- only WASP-18b is both more massive and dense while being
closer to its parent star. Figure~\ref{per_ecc_mass} demonstrates
this, plotting period vs. eccentricity for the know exoplanets with
periods less than 10 days. When examining this plot, another
characteristic of massive planets becomes apparent: they have a strong
tendency towards elliptical orbits -- only 3 of the 12
($\sim$25\%)transiting exoplanets that have masses greater than 2
$M_J$ and periods less than 10 days have $e=0$, not including those
planets with unknown eccentricity, while 3 more of these orbit stars
too faint to allow the orbital eccentricity to be measured readily. By
contrast, transiting planets with masses less than 2 $M_J$ and periods
less than 10 days have only a $\sim$21\% chance (13/63) of having a
non-zero eccentricity. While it is impossible to draw any definitive
conclusions with such a small sample size, these numbers suggest that
that more massive planets may in truth have longer periods and higher
eccentricities than less massive planets, although it is possible that
some of these non-zero eccentricities are artifacts arising from the
small number of RV measurements \citep{she2008}.

\begin{figure}[h]
\centering
\includegraphics[width=8.5cm]{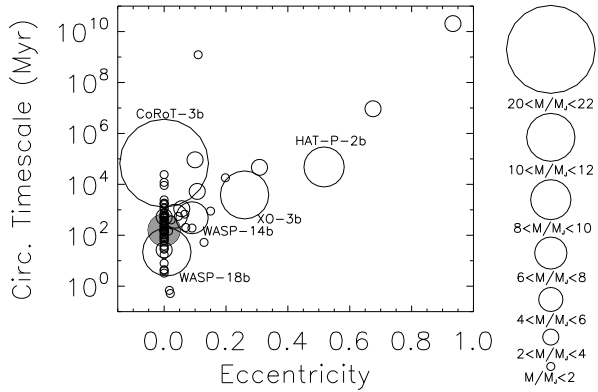}
\caption{Circularization timescale vs. eccentricity and planetary
  mass.  This figure shows the tidal circularization timescales of the
  known transiting exoplanets plotted again eccentricity assuming that
  the tidal quality factor $Q = 10^6$, with circle size indicative of
  the planetary mass. The tidal quality factor is a critical and
  technically unknown value -- short-period transiting exoplanets
  offer perhaps the best laboratory for increasing our understanding
  of this value. The circles along the bottom of the plot indicate the
  different planet mass ranges for each symbol size. CoRoT-14b is
  shaded gray and is otherwise readily identifiable by its large mass,
  zero eccentricity, and short circularization timescale: only
  WASP-18b is both more massive and has a shorter circularization
  timescale.  Notice that many of the more massive planets have
  relatively long circularization timescales and, unsurprisingly, tend
  to have eccentric orbits. While the numbers are small, there is a
  weak tendency for massive ($M>2M_J$) planets to have higher
  eccentricities at a given period. The notable exception of this
  hypothesis is the most massive object included in the plot:
  CoRoT-3b. This could be indicative of different formation mechanism
  than the smaller planets. Planetary parameters from the Exoplanet
  Encyclopedia (http://exoplanet.eu), except for CoRoT-14b.}
\label{circtime}
\end{figure}

An examination of the theory for tidal circularization and orbital
decay, arising from tides induced by the parent star on the exoplanet,
shows that this is not unexpected (see Figure~\ref{circtime}). Both of
these phenomena have timescales that go as $Q
M_pM_\star^{2/3}P^{13/3}R_p^{-5}$
\citep[see e.g.]{dob2004,ferr2008}. Assuming that $Q$, the quality factor, is
approximately equal for all gas giants, we would expect that high mass
planets with small radii will maintain their eccentricity (and
semi-major axis) longer -- a tendency further accentuated by the fact
that more massive planets have higher surface gravity, allowing them
to resist inflation caused in part by by incident radiation from the
parent star and therefore having smaller radii.

However, this does not explain the circular orbit of the high mass
planet/brown dwarf CoRoT-3b \citep{bou2008} -- its
circularization timescale is significantly longer than the age of the
universe. It is possible that CoRoT-3b might be eccentric -- the RV
observations used to measure this parameter are scattered over a year,
making it difficult to rule out small, non-zero eccentricities. If
both the adopted zero eccentricity and Q factor are correct, the
properties of CoRoT-3b would be indicative of {\it in situ} formation
rather than migration, the generally accepted process by which
short-period planets end up where they are. By contrast, CoRoT-14b is
less massive and closer to its host star, leading to a much shorter
circularization timescale. The observations of CoRoT14b are currently
consistent with a circularly orbit -- it would therefore come as no
surprise if this turns out to be the case in the end.

\begin{acknowledgements}
The team at the IAC acknowledges support by grant ESP2007-65480-C02-02
of the Spanish Ministerio de Ciencia e Inovaci\'{o}n.
M. Gillon acknowledges support from the Belgian Science Policy
Office in the form of a Return Grant.
Data presented herein were obtained at the W.M. Keck Observatory from
telescope time allocated to the National Aeronautics and Space
Administration through the agency's scientific partnership with the
California Institute of Technology and the University of
California. The Observatory was made possible by the generous
financial support of the W.M. Keck Foundation. 
The HIRES observations we obtained fell under the auspices of NASA's
key science program to support the CoRoT mission. 
The authors wish to recognize and acknowledge the very significant
cultural role and reverence that the summit of Mauna Kea has always
had within the indigenous Hawaiian community. We are most fortunate to
have the opportunity to conduct observations from this
mountain.
\end{acknowledgements}


\begin{figure}[h]
\centering
\includegraphics[width=9cm]{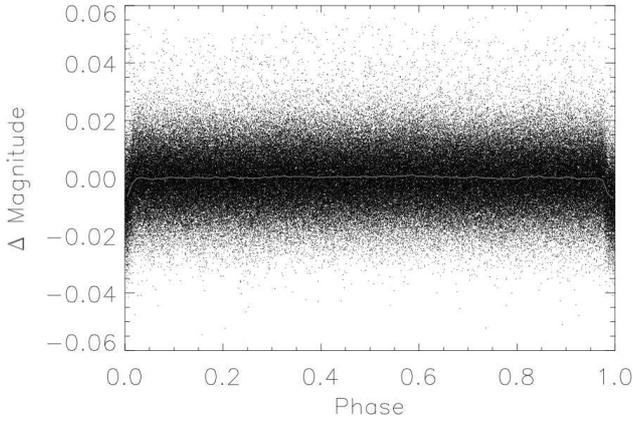}
\caption{The full phase-folded light curve of CoRoT-14. The light
  curve shown has been corrected for jitter, hot pixels, and other
  effects, then folded with the known period for CoRoT-14b, with the
  center of transit at 0. These observations were also binned into 100
  evenly-spaced bins, represented by a gray line. No out-of-transit
  variations are apparent. }
\label{photometry2}
\end{figure}

\end{document}